\documentclass[12pt,preprint]{aastex}

\begin{document}

\title{The Effect of the Approach to Gas Disk Gravitational Instability 
on the Rapid Formation of Gas Giant Planets}

\author{Alan P.~Boss}
\affil{Department of Terrestrial Magnetism, Carnegie Institution
for Science, 5241 Broad Branch Road, NW, Washington, DC 20015-1305}
\authoremail{aboss@carnegiescience.edu}

\begin{abstract}

 Observational evidence suggests that gas disk instability may be
responsible for the formation of at least some gas giant exoplanets,
particularly massive or distant gas giants. With regard to close-in 
gas giants, Boss (2017) used the $\beta$ cooling approximation to 
calculate hydrodynamical models of inner gas disk instability, finding 
that provided disks with low values of the initial minimum Toomre 
stability parameter (i.e., $Q_i < 2$ inside 20 au) form, fragmentation
into self-gravitating clumps could occur even for $\beta$ as high as 100
(i.e., extremely slow cooling).
Those results implied that the evolution of disks toward low $Q_i$ must 
be taken into account. This paper presents such models: initial disk 
masses of 0.091 $M_\odot$ extending from 4 to 20 au around a 1 $M_\odot$ 
protostar, with a range (1 to 100) of $\beta$ cooling parameters, the same
as in Boss (2017), but with all the disks starting with $Q_i = 2.7$, 
i.e., gravitationally stable, and allowed to cool from their 
initial outer disk temperature of 180 K to as low as 40 K. All the 
disks eventually fragment into at least one dense clump. The clumps 
were again replaced by virtual protoplanets (VPs) and the masses and 
orbits of the resulting ensemble of VPs compare favorably with those of 
Boss (2017), supporting the claim that disk instability can form gas 
giants rapidly inside 20 au, provided that sufficiently massive 
protoplanetary disks exist.

\end{abstract}


\section{Introduction}

 The spectacular success of the {\it Kepler} Mission (Borucki et al. 2010,
2011a,b) has revealed that super-Earth exoplanets are commonplace and
apparently considerably more abundant than gas giants, at least for short
period orbits. While core accretion appears to be the dominant formation 
mechanism for the exoplanets discovered to date, there is observational 
evidence suggesting that disk instability may also play a role, at least 
for the formation of massive or distant gas giants. As a result, hybrid
exoplanet population synthesis models (e.g., Boss 1998; Nayakshin 2010),
combining both core accretion (e.g., Mizuno 1980; Pollack et al. 1996) and 
disk instability (e.g., Cameron 1978; Boss 1997), may be needed to explain 
the plethora of exoplanetary system discoveries. Boss (2017) summarized 
the observations supporting such a role for disk instability in gas giant 
planet formation, including protoplanetary disk masses, temperatures, 
and lifetime estimates, as well as the demographics of mature exoplanets 
and evidence for large protoplanets embedded in protoplanetary disks. 

 Further observational evidence in support of disk instability has continued
to accumulate since Boss (2017), from both exoplanet demographics and
protoplanetary disk studies. Host stars with high metallicities presumably 
are accompanied by metal-rich protoplanetary disks, and hence are believed 
to favor the core accretion mechanism. Stellar metallicities have often been 
touted as the definitive test of giant planet formation mechanisms (e.g., 
Fischer \& Valenti 2005). However, recently analyses of stellar metallicities 
point to the need for a hybrid formation theory (e.g., Santos et al. 2017). 
Schlaufman (2018) showed that for solar-type stars, the distribution of 
transiting exoplanet masses as a function of stellar metallicity breaks 
up into two groups, with a break between about 4 $M_{Jup}$ and 10 $M_{Jup}$: 
exoplanets with masses below about 4 $M_{Jup}$ preferentially orbit 
metal-rich stars, while those with masses above 10 $M_{Jup}$ do not, 
implying a role for disk instability in forming the latter population. 
Narang et al. (2018) similarly found a break at 4 $M_{Jup}$, with host 
star metallicity dropping as the transiting exoplanet mass increases above 
this value. Goda \& Matsuo (2019) found that early-type stars can have 
massive exoplanets even for low metallicities. Maldonado et al. (2019) 
confirmed that host star metallicities drop as exoplanet masses increase, 
implying a role for disk instability for the more massive exoplanets.

 Many examples of protoplanetary disks with rings, or even grand spiral arms, 
have been imaged by the ALMA DSHARP survey (e.g., Huang et al. 2018), 
indicative of marginally gravitationally unstable disks, the rapid formation 
of massive planets that created and shaped the rings, or both processes. 
The Elias 2-24 disk appears to host several gas giant planets (Cieza et al. 
2017) at distances of 20, 52, and 87 au. The Elias 2-27 disk seems to 
contain a massive gas giant planet formed by disk instability (Meru et al. 
2017), though controversy continues over the source of the spiral arms 
(e.g., Forgan et al. 2018; Dong et al. 2018). The ongoing Subaru SEEDS 
survey has found nine distant companion candidates to date out of 68 young 
stellar objects studied (Uyama et al. 2017). Clarke et al. (2018) presented 
ALMA images arguing for the presence of four gas giant planets in the CI Tau 
disk, at an age of only 2 Myr, at orbital distances including $\sim$ 13, 39, 
and 100 au. The Herbig A5 star MWC 758 appears to be orbited by a dust
transition disk shepherded by an inner (35 au) 1.5-$M_{Jup}$ planet and 
by an outer (140 au) 5-$M_{Jup}$ planet (Baruteau et al. 2019).
Two accreting protoplanets have been imaged around the young 
star PDS 70 (Haffert et al. 2019), with masses in the range of 4 to 17
$M_{Jup}$ and 4 to 12 $M_{Jup}$, and orbital distances of $\sim$ 21 au
and $\sim 35$ au, respectively.
Core accretion has difficulties explaining the rapid formation
of gas giant planets at such distances, compared to disk instability
(e.g., Chambers 2006, 2016; Coleman \& Nelson 2016; cf. Boss 2011). Indeed, 
forming distant gas giants by disk instability has become conventional 
wisdom (e.g., Boley 2009; Nero \& Bjorkman 2009; Meru \& Bate 2010; 
Kratter \& Murray-Clay 2011; Rogers \& Wadsley 2012; Vorobyov 2013;
Madhusudhan et al. 2014; Rice et al. 2015; Young \& Clarke 2016;
Muller et al. 2018).

 This paper extends upon the work by Boss (2017), who used the $\beta$
cooling approximation for disk thermodynamics to study the fragmentation
process in disks with a wide range of $\beta$ values (1 to 100),
starting either close to gravitational instability (Toomre $Q_{min} 
= 1.3$) or gravitationally stable ($Q_{min} = 2.7$). Boss (2017)
found that fragmentation could occur even for high $\beta = 100$,
provided that the disk had a low initial value of $Q_{min} < 2$. The question
of disk fragmentation then becomes one of whether protoplanetary disks
can evolve into low $Q$, unstable configurations, or whether the
spiral arms that inevitably form first would transfer mass and angular momentum
fast enough to prevent fragmentation and dense clump formation,
as is commonly believed to be the case (e.g., Nixon et al. 2018).

 The purpose of the present paper is thus to examine the basic question of 
the approach to gas disk gravitational instability in disks which are 
cooling from an initially gravitationally stable configuration. 
The models start from the $Q_{min} = 2.7$ model of Boss (2017), with 
the same range of $\beta$ values, but with the minimum disk temperature 
constraints relaxed to 40 K (i.e., relaxed to that of the outer disk minimum
temperature for the unstable models with initial $Q_{min} = 1.3$). 
These models thus study the effect of varied cooling rates on the 
approach to a gravitationally unstable phase of disk evolution.

\section{Radiative Transfer and Beta Cooling Models}

 The inner regions of protoplanetary gas disks massive enough to become
gravitationally unstable are optically thick, requiring radiative
transfer to model the energy lost by disk radiation to the infalling
protostellar envelope. Boss (2001) calculated the first disk instability
models including radiation transfer in the diffusion approximation,
finding that fragmentation into dense clumps could occur in low $Q_{min}$
disks. Reviews by Durisen et al. (2007) and Helled et al. (2014) summarized 
the work on two- and three-dimensional hydrodynamical models of disk
instability, illustrating a variety of reasons for different results 
regarding fragmentation, ranging from the numerical grid resolutions of
finite difference codes and the smoothing lengths for smoothed-particle 
hydrodynamics (SPH) codes, to the accuracies of their gravitational 
potential solvers. The reviews concluded in part that the radiative transfer 
solver appeared to be critical to disk fragmentation, which requires
the disk to remain sufficiently cold for spiral arms to form self-gravitating,
contracting, dense clumps. Testing the radiative transfer codes against
analytical solutions has not resolved all the issues, e.g., Boley \& Durisen 
(2006) showed agreement of their cylindrical coordinate code with an analytical 
solution, while Boss (2009) found agreement with his spherical coordinate 
code with two other analytical radiative transfer solutions. 

 As a result of this impasse, and considering the heavy computational
overhead of three-dimensional radiative transfer, even in the
diffusion approximation, attention has turned to more simplified methods
for representing disk cooling processes. Gammie (2001) was the first to
propose that the outcome of gas disk gravitational instability depended 
primarily on the beta parameter $\beta = t_{cool} \Omega$, where $t_{cool}$ 
is the disk cooling time and $\Omega$ is the disk local angular velocity. 
When $\beta < 3$, Gammie (2001) suggested that the disk would fragment. 
As a result, a critical value of $\beta_{cr} = 3$ has become a standard for
predicting the outcome of the fragmentation process in protoplanetary disks.
However, subsequent work by many groups (as summarized in Boss 2017)
has questioned whether or not the value of $\beta_{cr} = 3$ is a correct
indicator for disk fragmentation, with other estimates of the true value ranging 
from $\beta_{cr} \sim 10$ to $\beta_{cr} \sim 30$, depending on the details
of the numerical code used, such as numerical resolution and the artificial
viscosity for SPH codes. More recently, Deng et al. (2017) found evidence
for $\beta_{cr} = 3$ for a novel meshless finite mass (MFM) code, but not for
a SPH code, the latter apparently a result of artificial viscosity.
Baehr et al. (2017) used local three-dimensional disk simulations to 
find $\beta_{cr} \sim 3$. Mercer et al. (2018) showed which of two 
approximate radiative transfer procedures is more accurate for protostellar
disks, and demonstrated that the effective value of $\beta$ in such
disks could vary from $\sim 0.1$ to $\sim 200$, i.e., a single, constant 
value of $\beta$ may not be capable of representing the full range
of physical conditions in gravitationally unstable disks.

\section{Numerical Methods and Initial Conditions}

 The numerical code is the same as that used by Boss (2017), which can be 
consulted for relevant details. The EDTONS code solves the three-dimensional 
equations of hydrodynamics and the Poisson equation for the gravitational 
potential, with second-order-accuracy in both space and time, on a spherical
coordinate grid (see Boss \& Myhill 1992). Note that an explicit artificial 
viscosity is not used in the models. Boss (2017) described how the $\beta$ 
cooling approximation was incorporated into the solution of the specific 
internal energy equation (Boss \& Myhill 1992), where the time 
rate of change of energy per unit volume, which is normally taken to be 
that due to the transfer of energy by radiation in the diffusion 
approximation, is defined in such a way that only cooling is permitted. 

 In the Boss (2017) models, the disk temperatures at given radial distances 
from the central protostar were not allowed to fall below their initial values,
which meant that initially warm and hot disks could not become any cooler 
than in their initial state, independent of the value of $\beta$.
Those disks could only approach gravitational instability by transporting 
disk mass inward so that the local gas disk surface density increased. 
This same minimum temperature constraint was imposed in all of the 
previous disk instability models in this series (see Boss 2017), and 
so was retained to allow comparisons with the earlier work. Here, though, 
we modify this constraint to allow disks to cool down to 40 K, regardless
of the initial radial temperature profile, which has a minimum of 180 K.
Figure 1 shows the initial temperature profile and minimum temperature
constraint for all the models. The profile decreases monotonically with
radial distance from the initial maximum of 600 K at 4 au.

 As in Boss (2017), the equations are solved on a grid with $N_r = 100$
or 200 uniformly spaced radial grid points, $N_\theta = 23$ theta grid points,
distributed from $\pi/2 \ge \theta \ge 0$, but compressed toward
the disk midplane, and $N_\phi = 512$ or 1024 uniformly spaced
azimuthal grid points. The radial grid extends from 4 to 20 au,
with disk gas flowing inside 4 au being added to the central 
protostar. The gravitational potential is obtained through a
spherical harmonic expansion, including terms up to $N_{Ylm} = 48$.
The $r$ and $\phi$ numerical resolutions are doubled when needed
to avoid violating the Jeans length (e.g., Boss et al. 2000) and Toomre
length criteria (Nelson 2006). If either of these two criteria is violated,
the calculation stops, and the spatial resolution is doubled in the 
relevant direction by dividing each cell into half while conserving mass and momentum. When dense clumps form, eventually the Jeans and Toomre length
criteria will be violated at their density maxima, even with $N_r = 200$ 
and $N_\phi = 1024$. The cell with the maximum density is then drained 
of 90\% of its mass and momentum, which is then inserted into a virtual protoplanet (VP, Boss 2005). These VPs orbit in the disk midplane, 
subject to the gravitational forces of the disk gas, the central protostar, 
and any other VPs, while the disk gas is subject to the gravity of the VPs. 
Those VPs that reach the the inner or outer boundaries are removed from the 
calculation. The VPs gain mass at the rate (Boss 2005, 2013) given by the 
Bondi-Hoyle-Lyttleton (BHL) formula (e.g., Ruffert \& Arnett 1994), as
well as the angular momentum of any accreted disk gas. The VPs are thus
handled identically to their treatment in Boss (2017).

  As in Boss (2017), the initial gas disk density distribution is that 
of an adiabatic, self-gravitating, thick disk in near-Keplerian rotation 
about a stellar mass $M_s$ (Boss 1993), with an initial midplane density 
chosen to enforce near-Keplerian rotation. The inner disk radius is 4 au,
and the outer disk radius is 20 au. The initial disk mass $M_d$ is then
0.091 $M_\odot$, and the initial protostellar mass is $M_s = 1.0 M_\odot$. 
The initial outer disk temperature is set to 180 K for all models,
so that all begin from a gravitationally stable configuration with
an initial minimum value of the Toomre (1964) $Q$ gravitational 
stability parameter of 2.7. Only the value of $\beta$ was varied
in the models (Table 1), with the same eight values used as in Boss (2017):
1, 3, 10, 20, 30, 40, 50, and 100.

 Cieza et al. (2018) used ALMA to show that FU Orionis disks have radii 
less than 20 to 40 au and masses of ∼ 0.08 to 0.6 $M_\odot$. Such disks 
are the observational analogs of the marginally gravitationally unstable
disks to be studied here.

\section{Results}

  Table 1 lists the basic results for all of the models:
the final times reached, the final and maximum number of virtual 
planets ($N_{VP}$) that formed, and the amount of disk mass lost
by accretion onto either the VPs or the central protostar, in Jupiter
mass units. The final times reached ranged from 288 yrs to 462 yrs,
similar to the results of Boss (2017). The initial orbital period 
of the disk at the inner edge is 8.0 yr and 91 yrs at the outer edge, 
so clearly the models spanned a time period long enough for many
revolutions in the inner disk, and multiple revolutions in the outer 
disk. Each model required over one year of time to compute, each running 
on a separate, single core of the Carnegie memex cluster at Stanford 
University.

\subsection{Model be5}

 Figures 2 and 3 present the results for model be5, which is typical of
the results for all of the models here. Figure 2 shows that the initially
nearly axisymmetric gas disk begins to form strong spiral arms just
outside the hot inner disk, and these arms continue to grow and extend
to the outer disk. By the end of the evolution, the arms are dominated 
by $m = 1$ one-armed spirals, with maximum amplitude $a_{max} \approx 1.4$, 
along with $m = 2$ two-armed spirals with $a_{max} \approx 0.5$ and 
$m = 3$ three-armed spirals with $a_{max} \approx 0.3$. The caption
for Figure 2 lists the maximum midplane densities reached, which started
at $1.0 \times 10^{-10}$ g cm$^{-3}$, and rose as high as
$2.6 \times 10^{-9}$ g cm$^{-3}$ after 162 yrs for the densest clump 
seen in Figure 2b. These clump densities should be compared to the
relevant free fall times when considering whether the clumps might 
contract to even higher densities and survive as VPs. The free 
fall time ($t_{ff}$) for a pressure-less, uniform density ($\rho$)
gas sphere is $t_{ff} = (3 \pi / 32 G \rho)^{1/2}$, where $G$ is the
gravitational constant. For $\rho = 10^{-10}$ g cm$^{-3}$, $t_{ff} = 6.7$ yrs, 
while for $\rho = 10^{-9}$ g cm$^{-3}$, $t_{ff} = 2.1$ yrs. Clearly
the clumps formed in model be5 reach maximum densities high enough to
plausibly contract to higher densities on time scales considerably less
than a single orbital period at their orbital distances, which are 
about 12 au for Figure 2b, where the orbital period is $\sim 40$ yrs.

 Figure 3 shows the midplane temperature evolution, starting from 180 K
outside the hot inner disk, and cooling down to as low as 40 K in the
outer disk after $\sim 162$ yrs. For model be5, with $\beta = 30$, 
$t_{cool} = \beta/ \Omega = \beta P / (2 \pi) \approx  4.8 P$, where
$P$ is the local disk rotation period. Given the initial disk rotation 
periods of 8.0 yrs at the inner edge and 91 yrs at the outer edge, the 
nominal cooling times for model be5 are then about 38 yrs at the inner 
edge and 436 yrs at the outer edge. The cooling time is coded (Boss 2017)
as being the ratio of the specific internal energy to the time rate of
change of the specific internal energy. With that definition, and in the
absence of compressional or other heating terms, the specific internal
energy could formally drop to zero in one cooling time. Hence the rate of
inner and outer disk cooling seen in the models can be understood in the
manner that $\beta$ cooling was defined in both the present and 2017 models.
Note that the innermost disk forms high temperature spiral features on
top of the hot inner disk minimum temperature constraint, though the inner 
gas disk remains relatively featureless in Figure 2, compared to the outer
disk.

\subsection{VPs Formation and Evolution}

 The maximum midplane density in model be5 in Figure 2b occurs in the 
clump located close to 12 noon at an orbital radius of $\sim 10$ au. The disk 
gas has a maximum temperature of $\approx 100$ K at the maximum density of 
$2.6 \times 10^{-9}$ g cm$^{-3}$ at 162 yrs, yielding a Jeans mass
of $\approx 1.2 \times 10^{30}$ g, or about 0.6 $M_{Jup}$. The mass of 
this densest clump is $\approx 1.3 \times 10^{30}$ g, or about 0.65 
$M_{Jup}$, making the clump marginally gravitationally bound. Shortly
after 162 yrs, the clump maximum density exceeds the Toomre criterion,
and as the model had already been refined to the highest spatial resolution,
a VP was inserted in the center of the clump with an initial mass of
$\sim 0.001 M_{Jup}$, 90\% of the mass within the cell with the maximum
midplane density. The VP thereafter orbits the central protostar
and rapidly accretes disk gas and orbital angular momentum, as shown
in Figure 4, at the Bondi-Hoyle-Littleton rate (Ruffert \& Arnett 1994).
Figure 4 displays the masses of all the VPs formed during the evolution
of model be5, which at one time (Table 1) had as many as five VPs in 
orbit, though only a single VP was still active at the end of the 
evolution at 422 yrs. The others hit either the inner (4 au) or outer
(20 au) grid boundary, and were thereafter dropped from consideration.
Note that in a more realistic disk model, VPs that encounter these
artificial orbital limits need not be lost altogether, as they may very
well orbit for significant periods without being accreted by the
central protostar, or ejected from the system by mutual close encounters.

 Figure 4 shows that some of the VPs in model be5 gained mass at the rate of 
about 1 $M_{Jup}$ per 100 yrs, or $\sim 1 \times 10^{-2} M_{Jup}$ yr$^{-1}$.
This rate is considerably higher than estimated mass accretion rates
for observed accreting young companions embedded in circumstellar disks,
where objects with masses of $\sim$ 10 $M_{Jup}$ accrete no faster
than $\sim 1 \times 10^{-6} M_{Jup}$ yr$^{-1}$ (e.g., Cugno et al. 2019).
However, the very fact that these observational estimates were based on 
direct imaging of young companions implies that these circumstellar disks are 
not as optically thick, and hence not as massive, as the protoplanetary disks 
modeled here, as well as being objects at much later phases of evolution 
than the newly-formed VPs studied here. Model be5 formed VPs with a
total mass of $\sim 4 M_{Jup}$ by the end of the calculation. Table 1 
shows that the model be5 disk lost $4.5 M_{Jup}$ during the evolution,
so most of this mass loss went to VP formation, and only $\sim 0.5 M_{Jup}$
was accreted by the central protostar. The central protostar mass
accretion rate was thus $\sim 10^{-6} M_\odot$ yr$^{-1}$, a typical
rate for a marginally gravitationally unstable disk. In comparison,
T Tauri stars and FU Orionis outburst stars have mass accretion rates 
of $\sim 10^{-7} M_\odot$ yr$^{-1}$ and $\sim 10^{-4} M_\odot$ yr$^{-1}$,
respectively (e.g., Hartmann \& Kenyon 1996).

 Figure 5 presents the results for the VPs formed in all eight models listed
in Table 1, showing similar behavior as in model be5: many VPs form, but only
a fraction survive long enough to accrete sufficient gas to reach gas
giant planet masses. Figure 5 is quite similar to the corresponding Figure 8a
in Boss (2017), showing the results for the varied $\beta$ cooling
models where the midplane temperatures were not allowed to drop below
the initial values, for the model with the lowest outer disk initial 
temperature (40 K, compared to 180 K here). The results suggest
that disk instability can form dense clumps in either case: starting
with a relatively cold disk with a range of $\beta$ values (Boss 2017),
or starting with a relatively warm disk and waiting for $\beta$ cooling
to increase the degree of gravitational instability of the disks.

 Figure 6 shows the results for all eight models in terms of the evolution
of the orbital radii of the VPs that form. Formation occurs in the disk
sweet spot between about 6 au and 10 au, where the midplane temperatures are
not as high as in the inner disk (e.g., Figure 3), the midplane densities
are high, and the orbital period ($\sim 20$ yr) is shorter than in
the outer disk ($\sim 90$ yr). While a significant fraction manage to continue
to orbit in the sweet spot, 73 VPs reach the inner boundary at 4 au, while
another 27 strike the outer boundary, as noted in Table 1. The VPs migrate
as a result of the gravitational torques they receive from the evolving
disk spiral arms (e.g., Figure 2), as well as from close encounters with
each other. Again, this behavior is quite similar to that of the corresponding
models in Boss (2017, Figure 8b). The circled points in Figure 6 show
the final locations of the VPs, which are distributed more or less
uniformly throughout the disk.

 Figure 7 shows the masses and orbital radii of all of the VPs from the
present set of models, sampled periodically throughout the evolutions,
and plotted as in Figure 9 of Boss (2017). For comparison, Figure 8 shows
the known exoplanets as of July 10, 2019, from the Extrasolar Planets 
Encyclopedia (exoplanets.eu). Table 2 shows the percentage of
the total number of exoplanets in Figures 7 and 8 with masses above 
$\approx 0.1 M_{Jup}$ broken up into separation bins of 2 au. 
It will be seen that the two distributions look reasonably similar for 
semi-major axes less than about 8 au, as was the case for the Boss (2017) 
models. Note also that the present models sample
only the effects of varying the disk cooling rate, just one parameter 
out of the many that may need to be varied to reproduce exoplanet
demographics (e.g., disk and stellar masses). The results shown in Figure 7 
are also broadly consistent with the results of the disk instability
models by Forgan et al. (2015), though a direct comparison is not
appropriate, as they considered disks with radii of 100 au, five times
larger than the present models. 

 Finally, Table 1 demonstrates that though all the models started with
identical initial conditions, the models with the lower $\beta$ values
fragmented into considerably larger numbers of VPs than the higher
$\beta$ models, as expected. They also lost considerably more VPs to
migrations into the grid boundaries, consistent with their greater numbers
formed and more vigorous spiral arms, which are the principal agent
of VP orbital migration, as the mass in any individual spiral arm is
typically greater than that of the most massive VPs that form.
Nevertheless, it is remarkable that even the models with the highest
values of $\beta$ still fragmented into at least one dense clump that
required the formation of a VP, and that in each model, at least one VP
remained active at the end of the calculations. Allowing the
disks to slowly cool down to 40 K was critical for obtaining spiral
arm formation and eventual dense clump formation to occur in the
models with large values of $\beta$; the Boss (2017) models with 
large $\beta$ values did not fragment when the disk was prevented
from cooling to such a low temperature. Midplane temperatures of
40 K at 10 au are consistent with spectral energy distributions for
T-Tauri-star disks (Lachaume et al. 2003; D’Alessio et al. 2006).

 The results imply that
the choice of the value of $\beta$ for disk fragmentation may not be
as critical as has been generally assumed or found to be the case.
As noted in the Introduction and summarized in Boss (2017), many
papers have questioned whether the Gammie (2001) value of $\beta_{cr} = 3$ 
is a universal criterion, given that other estimates range from $\sim 10$ 
to $\sim 30$ and depend on the details of the numerical code used. E.g., 
Gammie (2001) used a razor-thin (2D) numerical model in the shearing-sheet
approximation, which can only represent a small region of a large-scale
disk. Considering that the clumps formed in the present models result
from the interactions of large-scale spiral arms (Figure 2), it is hard to
see how a local analysis could be relevant for such large-scale fragmentation.
While Rice et al. (2003) used 3D SPH disk models to confirm Gammie's basic
claim with a higher value of $\beta_{cr} \sim 6$, their models were 
limited to a rather low resolution of 250,000 ($\sim 64^3$) particles, 
limiting the code's ability to achieve convergence. In contrast, the present
models employ $N_\phi = 512$ and 1024 in the coordinate critical for
the growth of non-axisymmetric structures. Meru \& Bate (2011b) found 
that convergence was not achieved for SPH models with as many as 16,000,000
particles, finding that fragmentation occurred for larger $\beta$ values
as the resolution was increased, leading them to doubt whether a critical 
value of $\beta$ exists. Meru \& Bate (2011a) found that estimates of
$\beta_{cr}$ also depend on the assumed underlying disk density profile,
again arguing against the generality of a specific value for $\beta_{cr}$.
Further comments about previous $\beta$-cooling models may be found in 
Boss (2017).

Boss (2017) summarized the state of radiative transfer models of
disk instability, which should yield a more correct handling of disk
thermodynamics than simple $\beta$ cooling models. As noted in the
reviews by Durisen et al. (2007) and Helled et al. (2014), the outcome
of radiation hydrodynamics (RHD) disk instability models appears to be
dependent on a range of numerical factors. Even when RHD codes are
tested against analytical solutions, a clear decision need not result
(e.g., cf. Boley \& Durisen 2006, Boss 2009). Boley (2009) demonstrated
that large-scale fragmentation is possible with his RHD code, 
using an azimuthal resolution of $N_\phi = 512$, when 
initially isothermal disks with radii of $\sim$ 300 au are formed from gas
infalling beyond 60 au from the central protostar. No fragmentation
inside $\sim 20$ au was observed, as there was negligible disk mass
in the inner regions, by design of the numerical experiment.
Stamatellos \& Whitworth (2008) presented two SPH RHD models of disks
with radii of 40 au that did not undergo fragmentation. However,
their two models were limited to 200,000 particles representing the
gas disk, which appears to be insufficient for achieving convergence.
Finally, Mercer et al. (2018) found that their RHD results implied 
that the effective value of $\beta$ varied throughout the disk, from 
$\sim 0.1$ to $\sim 200$, an even wider range than studied in the 
present set of models.

\section{Conclusions}

 These models are intended to be first steps toward
creating a hybrid model for exoplanet population synthesis, where a
combination of core accretion and disk instability works in tandem to try 
to reproduce the exoplanet demographics emerging from numerous large surveys 
using ground-based Doppler spectroscopy and gravitational microlensing 
or space-based transit photometry (e.g., {\it Kepler}, TESS).

 The Boss (2017) models showed that the outcome of a phase of disk
gravitational instability depends more strongly on the initial conditions 
adopted for the models than on the assumed disk cooling rate $\beta$.
The present models have studied the evolution of protoplanetary disks 
into gravitationally unstable configurations, which is evidently just
as important a factor as the disk cooling process. Remarkably,
the models have shown that starting from a gravitationally stable,
high Toomre $Q$ disk, disks with a large range of cooling rates,
from $\beta$ = 1 to 100 (Table 1), eventually become gravitationally 
unstable, form numerous spiral arms, and then dense clumps requiring
the insertion of VPs representing newly formed gas giant protoplanets.
In combination with the results of Boss (2017), the models imply that
protoplanetary disks with masses of $\sim 0.1 M_\odot$ (e.g., FU Orionis
star disks) should be able to form gas giants in the region from 
$\sim$ 4 au to $\sim$ 20 au. This implies the existence of a largely unseen 
population of gas giants orbiting solar-type stars, which could be detected 
by the gravitational microlensing survey and coronagraphic direct 
imaging technology efforts of the NASA WFIRST space mission, slated
for launch around 2025.
 
 In order to continue to study the viability of the disk instability
mechanism for giant planet formation, the author is currently computing
several other suites of protoplanetary disk models. One set is a 
continuation of the $\beta$ cooling models presented here, but with
a significant further enhancement of the highest spatial resolution 
permitted before the VP artifice is employed. While the present models
were restricted to a maximum of 200 radial and 1024 azimuthal grid
points before VP insertion, the suite currently running doubles both
of these maxima, to 400 in radius and 2048 in azimuth. These are the
highest spatial resolution grid models run to date with the EDTONS code, 
but at a penalty of running about eight times slower than the highest 
resolution models presented in this paper (time step halved for four 
times as many grid points). The second set was noted previously by 
Boss (2017), and these are flux-limited diffusion approximation (FLDA)
models, starting with the same initial conditions as the present 
suite of models. The FLDA models run considerably slower than 
$\beta$ cooling models for the same grid resolution, and hence
the FLDA models are still in progress. The goal of this second suite
is to compare the FLDA results with those of the $\beta$ cooling models,
at the same spatial resolution and same initial conditions, to attempt
to determine if there is an optimal value of $\beta$ that could be used 
in future disk instability models, thereby avoiding the computational
burden associated with the FLDA approach.

 The referee provided a number of helpful suggestions for improving 
the manuscript. The calculations were performed on the Carnegie memex
cluster at Stanford University. I thank Michael Acierno and Floyd Fayton 
for their able assistance with the use of memex.

\clearpage
\begin{deluxetable}{lccccccc}
\tablecaption{Initial conditions and results for the models,
including the number of VPs lost to the inner and outer
boundaries ($\delta N_{VPi}$ and $\delta N_{VPo}$, respectively),
and the total amount of disk mass lost to the central protostar or to
VP formation and accretion ($\delta M_d$, in $M_{Jup}$ units).}
\label{tbl-1}
\tablewidth{0pt}
\tablehead{\colhead{Model} 
& \colhead{$\beta$}
& \colhead{final time (yrs)} 
& \colhead{final $N_{VP}$} 
& \colhead{maximum $N_{VP}$} 
& \colhead{$\delta N_{VPi}$} 
& \colhead{$\delta N_{VPo}$} 
& \colhead{$\delta M_d$} }
\startdata

be1   &    1   &  288. &   4  &  8  & 33 & 10 & 4.5 \\                              

be2   &    3   &  407. &   3  &  7  & 28 & 13 & 3.5 \\

be3   &   10   &  391. &   5  &  5  & 2  & 3  & 4.0 \\
 
be4   &   20   &  447. &   3  &  3  & 2  & 0  & 1.5 \\

be5   &   30   &  422. &   1  &  5  & 5  & 1  & 4.5 \\

be6   &   40   &  462. &   1  &  2  & 1  & 0  & 1.0 \\

be7   &   50   &  450. &   2  &  3  & 1  & 0  & 1.5 \\
 
be8   &  100   &  417. &   1  &  2  & 1  & 0  & 2.5 \\
   
\enddata
\end{deluxetable}

\clearpage
\begin{deluxetable}{lcc}
\tablecaption{Percentage of the total number of exoplanets in the
present models (Figure 7) and from the Extrasolar Planets Encyclopedia
observations (Figure 8) with masses above $0.1 M_{Jup}$, sorted into 
separation bins of 2 au. The distributions are not inconsistent,
considering the numerical bias for exoplanets near the finite radius outer
edge of the numerical grid (20 au) and the observational bias against
detections of long period orbits by transits or Doppler spectroscopy.}
\label{tbl-2}
\tablewidth{0pt}
\tablehead{\colhead{separation (au)} 
& \colhead{models (\%)}
& \colhead{observations (\%)}}
\startdata

4 - 6   & 28. & 50.  \\                              
6 - 8   & 31. & 32.  \\
8 - 10  & 19. & 9.   \\
10 - 12 & 7.  & 4.5  \\
12 - 14 & 6.  & 0.   \\
14 - 16 & 1.5 & 4.5  \\
16 - 18 & 1.5 & 0.   \\
18 - 20 & 6.  & 0.   \\
   
\enddata
\end{deluxetable}

\clearpage

\begin{figure}
\vspace{-2.0in}
\plotone{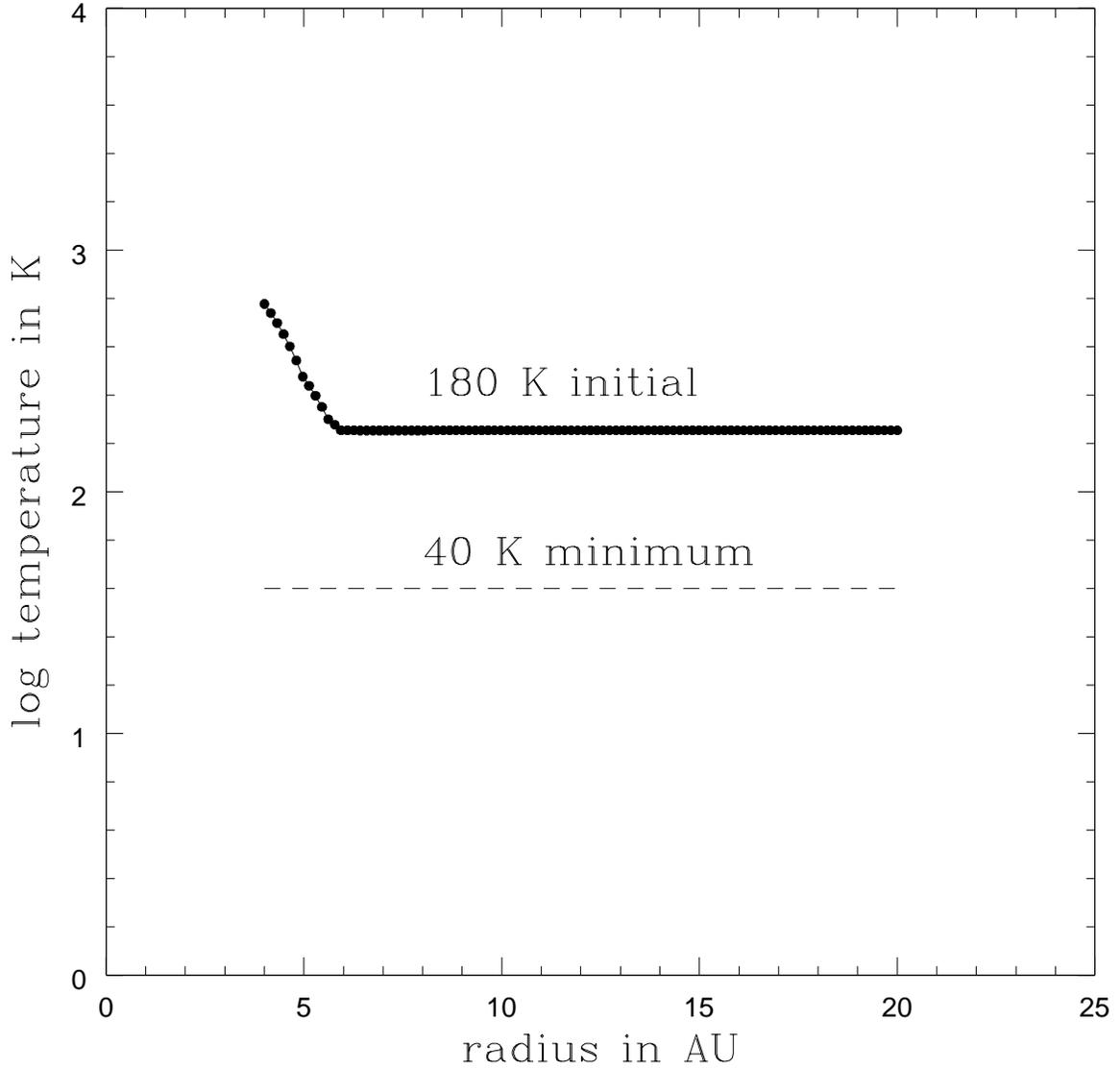}
\caption{Initial midplane temperature profile for all the models,
starting with an outer disk temperature of 180 K, leading to Toomre
$Q = 2.7$ in the initially gravitational stable outer disk. The minimum
outer disk temperature of 40 K, imposed during the evolutions, is
also indicated.}
\end{figure}

\begin{figure}
\vspace{-1.0in}
\plotone{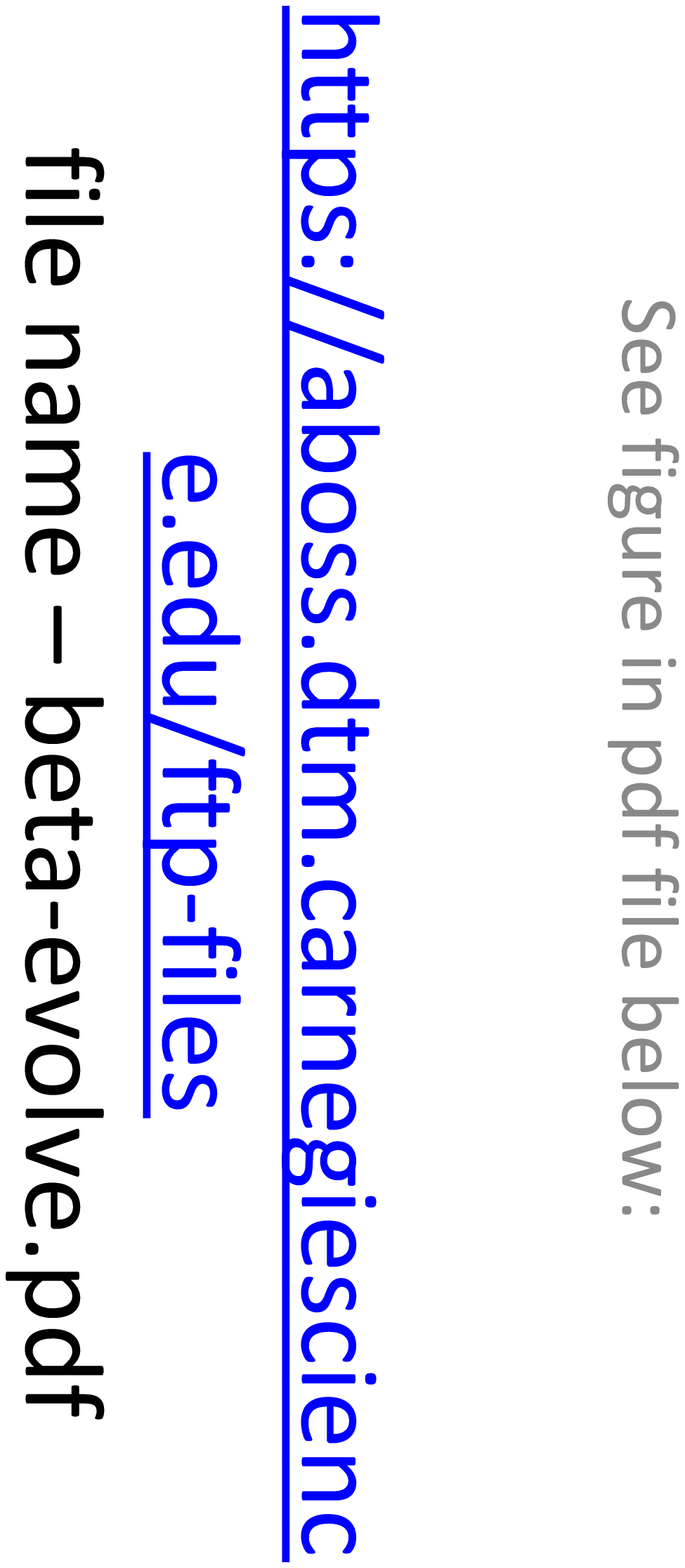}
\caption{Equatorial (midplane) density contours for model be5 after 
(a) 38.5 yr, (b) 162 yr, (c) 318 yr, and (d) 422 yr. The disk has an inner 
radius of 4 au and an outer radius of 20 au. Contours are labelled in 
log cgs units. Maximum midplane gas densities at each time are:
(a) $1.2 \times 10^{-10}$ g cm$^{-3}$; (b) $2.6 \times 10^{-9}$ g cm$^{-3}$;
(c) $1.3 \times 10^{-9}$ g cm$^{-3}$; and (d) $5.4 \times 10^{-10}$ g cm$^{-3}$. 
The initial maximum midplane density is $1.0 \times 10^{-10}$ g cm$^{-3}$.}
\end{figure}

\begin{figure}
\vspace{-1.0in}
\plotone{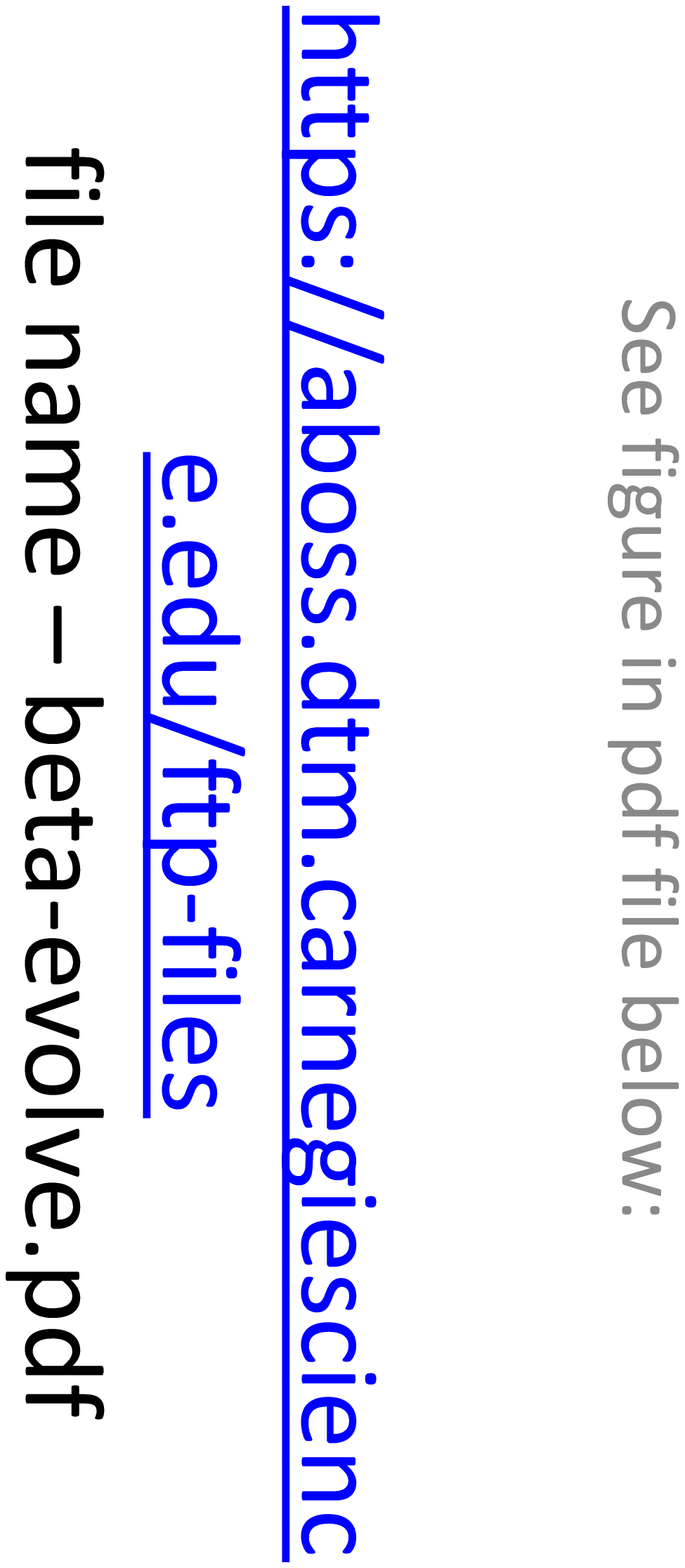}
\caption{Equatorial (midplane) temperature contours for model be5 after 
(a) 38.5 yr, (b) 162 yr, (c) 318 yr, and (d) 422 yr, plotted as in Figure 2. 
Contours are labelled in log K units. The model starts with an  initial 
minimum temperatures of 180 K (light orange color) and can cool down to
a minimum temperature of 40 K (light green color).}
\end{figure}

\begin{figure}
\vspace{-2.0in}
\plotone{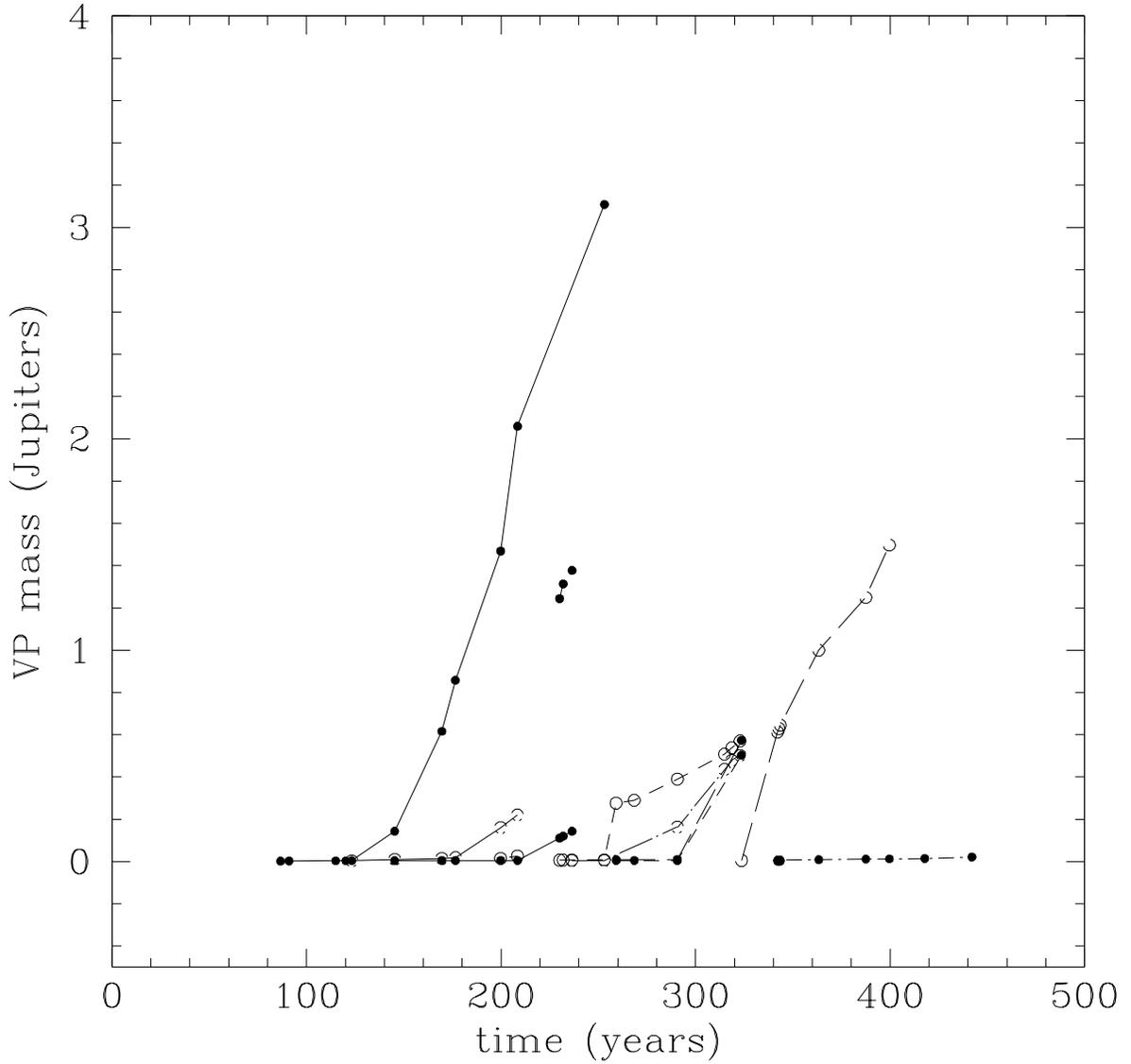}
\caption{VP mass as a function of time for all of the VPs 
formed by model be5, with $\beta = 30$, sampled every 40,000 time 
steps throughout their evolutions. The VP masses increase by BHL accretion
from the disk gas, while VPs that hit the inner (4 au) or outer (20 au)
boundaries are removed from the evolutions. Differing symbols and lines
are used to help delineate different VPs.}
\end{figure}

\begin{figure}
\vspace{-2.0in}
\plotone{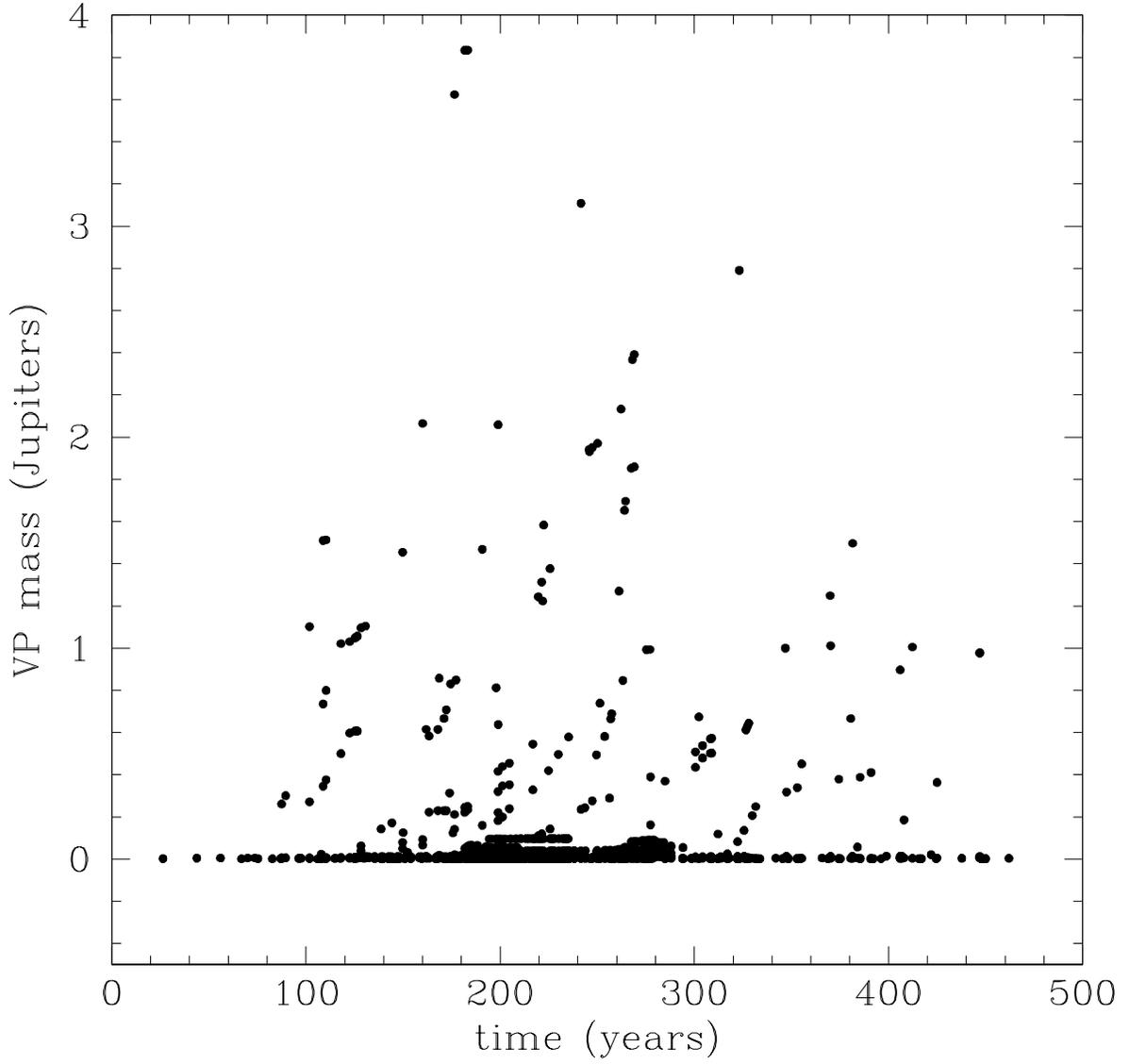}
\caption{VP mass as a function of time for all of the VPs 
formed by all eight models in Table 1, plotted as in Figure 4.}
\end{figure}

\begin{figure}
\vspace{-2.0in}
\plotone{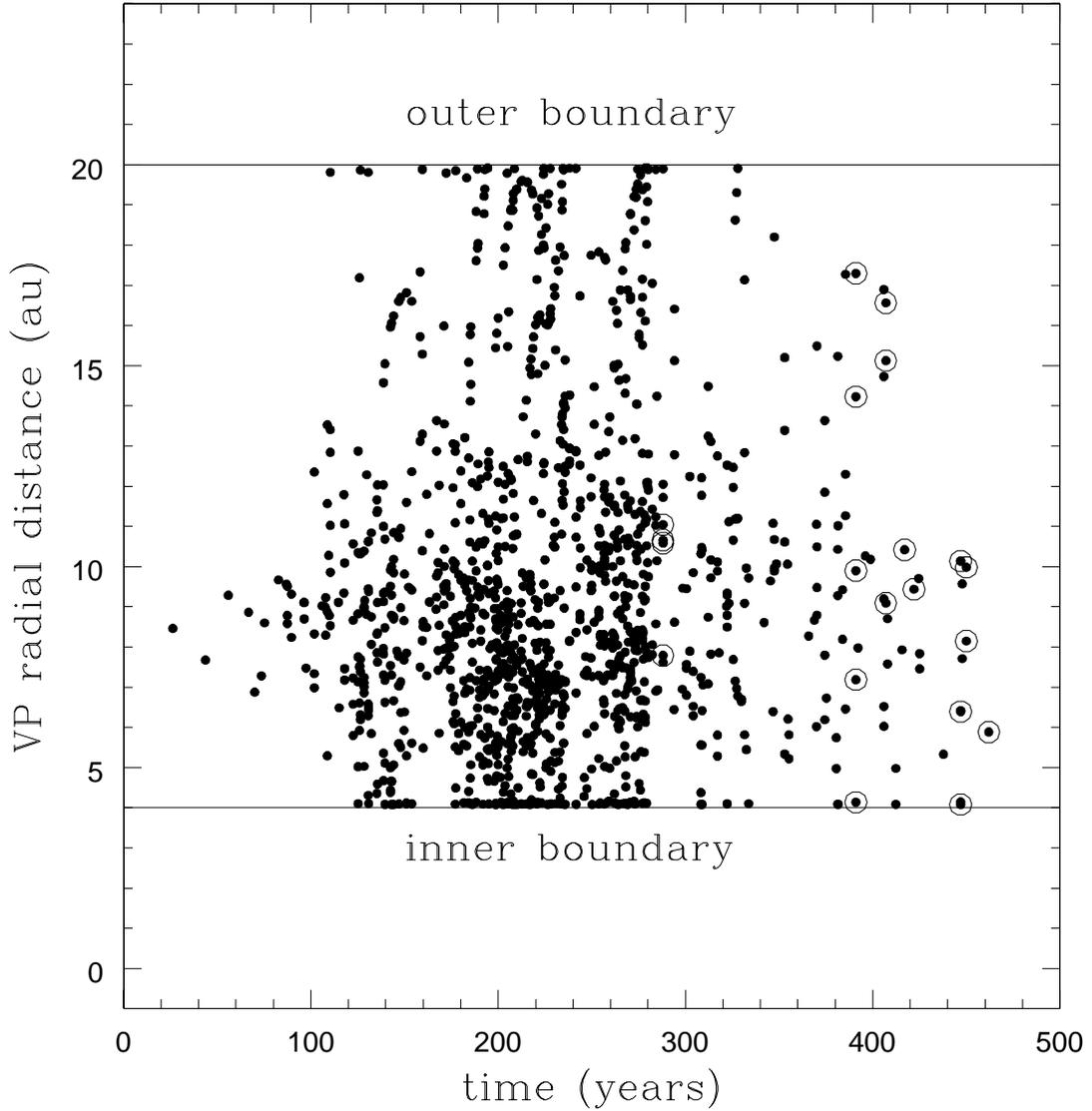}
\caption{VP orbital radii as a function of time for all of the VPs 
formed by the eight models in Table 1, plotted as in Figure 4. The orbital radii both 
increase and decrease by disk migration and close encounters after forming in the 
$\sim$ 6 au to 10 au region. VPs that hit the inner or outer boundaries are removed.
Circled points designate the VP locations at culmination of each evolution.}
\end{figure}

\begin{figure}
\vspace{-2.0in}
\plotone{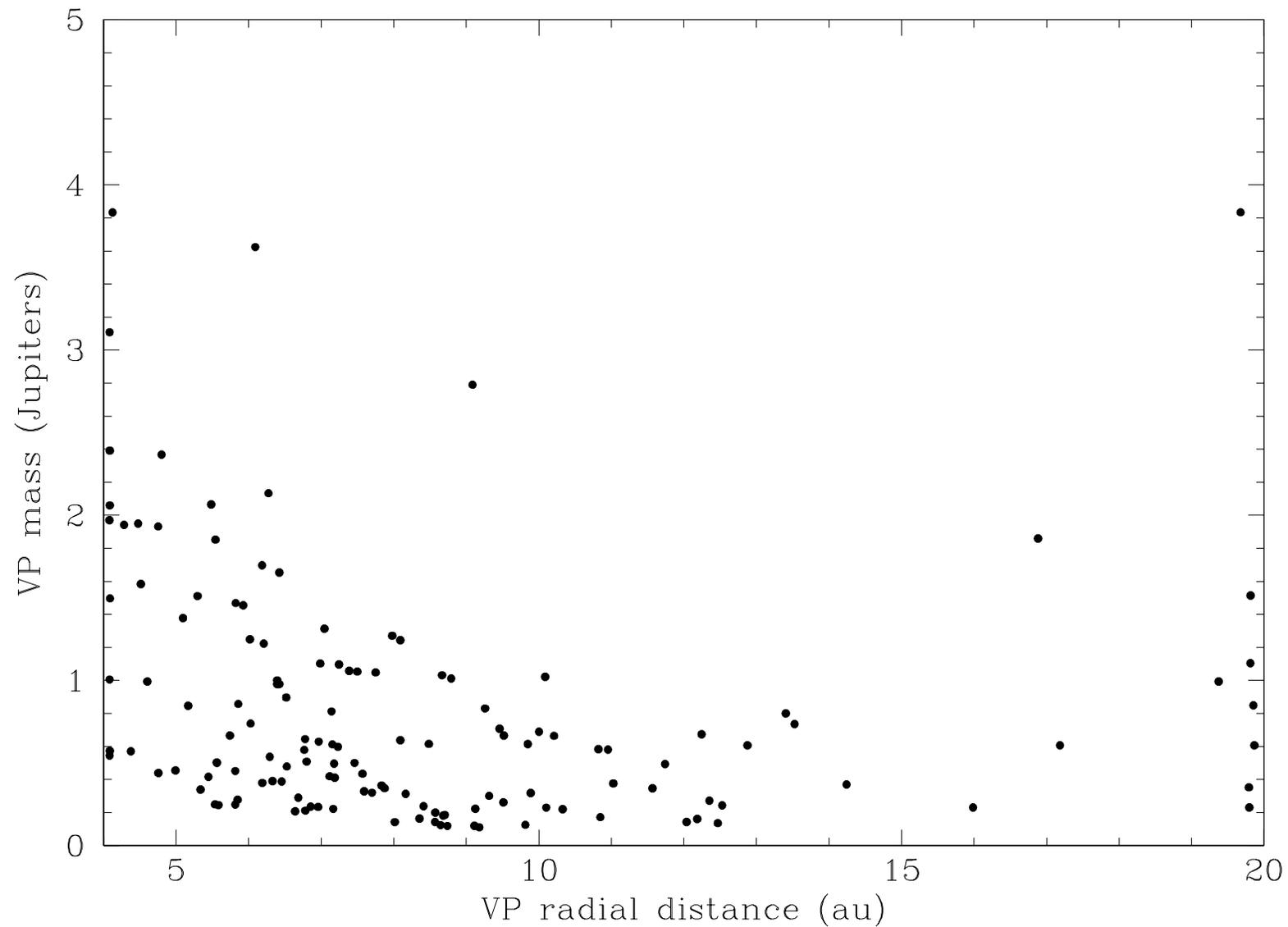}
\caption{Masses and orbital radii of all of the VPs formed by the eight models in
Table 1, sampled about 40 times throughout their evolutions, as in Boss (2017,
Figure 9). The region plotted extends from 4 au to 20 au with masses between 0.1 
and 5 $M_{Jup}$, for comparison (Table 2) with the observed exoplanets in Figure 8.}
\end{figure}

\begin{figure}
\vspace{-2.0in}
\plotone{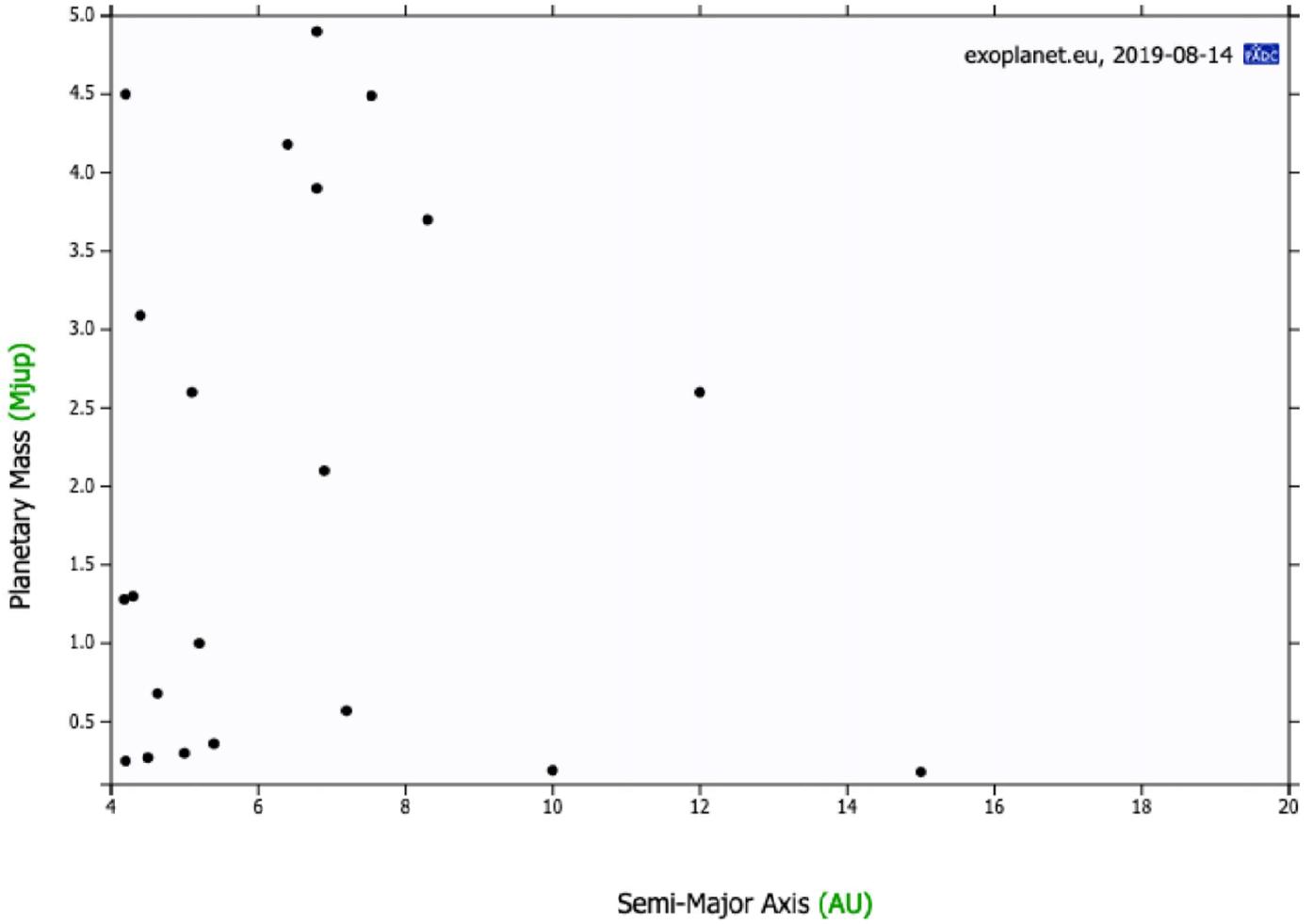}
\caption{All exoplanets in the Extrasolar Planets Encyclopedia 
(exoplanets.eu) as of August 14, 2019, for masses between 0.1 and 5 $M_{Jup}$ 
and semi-major axes between 4 au and 20 au, for comparison (Table 2) with the 
model results in Figure 7.}
\end{figure}

\end{document}